\pdfoutput=1

\documentclass[11pt,twoside,a4paper,cmspaper,final,collab]{cms-tdr}
\def\svnVersion{365921}\def\cmsCernNoTag{CERN-PH-EP/2015-210}\def\cmsCernDate{\today}\def\cmsMessage{Published in Physics Letters B as \href{http://dx.doi.org/10.1016/j.physletb.2016.06.027}{\doi{10.1016/j.physletb.2016.06.027.}}}

\begin{document}\cmsNoteHeader{FSQ-13-006}

\hyphenation{had-ron-i-za-tion}
\hyphenation{cal-or-i-me-ter}
\hyphenation{de-vices}

\RCS$Revision: 346724 $
\RCS$HeadURL: svn+ssh://svn.cern.ch/reps/tdr2/papers/FSQ-13-006/trunk/FSQ-13-006.tex $
\RCS$Id: FSQ-13-006.tex 346724 2016-06-11 15:17:34Z alverson $
\newlength\cmsFigWidth
\ifthenelse{\boolean{cms@external}}{\setlength\cmsFigWidth{0.99\columnwidth}}{\setlength\cmsFigWidth{0.6\textwidth}}
\ifthenelse{\boolean{cms@external}}{\providecommand{\cmsLeft}{top\xspace}}{\providecommand{\cmsLeft}{left\xspace}}
\ifthenelse{\boolean{cms@external}}{\providecommand{\cmsRight}{bottom\xspace}}{\providecommand{\cmsRight}{right\xspace}}
\ifthenelse{\boolean{cms@external}}{\providecommand{\suppMaterial}{\cite{suppl1}\xspace}}{\providecommand{\suppMaterial}{App.~\ref{app:suppMat}\xspace}}
\providecommand{\NA}{\text{---}\xspace}
\providecommand{\HIJINGWITHVER} {\textsc{hijing}\,{1.383}\xspace}
\providecommand{\EPOS} {\textsc{epos}\xspace}
\providecommand{\EPOSWITHVER} {\textsc{epos-lhc}\xspace}
\providecommand{\DIPSY} {\textsc{dipsy}\xspace}
\providecommand{\QGSJETWITHVER} {\textsc{qgsjetii}--{04}\xspace}
\providecommand{\QGSJET} {\textsc{qgsjetii}\xspace}
\providecommand{\STARLIGHT} {\textsc{starlight}\xspace}
\providecommand{\DPMJETWITHVER} {\textsc{dpmjet}\,{3.06}\xspace}
\providecommand{\DPMJETWITHVERSL} {\textsc{dpmjet}\,{3.05}\xspace}
\providecommand{\DPMJET} {\textsc{dpmjet}\xspace}
\providecommand{\PYTHIAWITHVERSL} {\textsc{pythia}\,{6.4.26}\xspace}

\providecommand{\sigmainel} {\ensuremath{\sigma_{\text{inel}}}\xspace}
\providecommand{\sigmaobs} {\ensuremath{\sigma_{\text{obs}}}\xspace}
\providecommand{\sigmahadlvl} {\ensuremath{\sigma_{\text{vis}}}\xspace}
\providecommand{\nobs} {\ensuremath{N_{\text{obs}}}\xspace}
\providecommand{\nhad} {\ensuremath{N_{\text{had}}}\xspace}
\providecommand{\nhadnocut} {\ensuremath{N_{\text{inel}}}\xspace}
\providecommand{\nem} {\ensuremath{N_{\gamma\Pp}}\xspace}
\providecommand{\nnoise} {\ensuremath{N_{\text{noise}}}\xspace}
\providecommand{\ehf} {\ensuremath{E_{\mathrm{HF}}}\xspace}
\providecommand{\chad} {\ensuremath{c_{\text{vis}}}\xspace}

\cmsNoteHeader{FSQ-13-006}
\title{Measurement of the inelastic cross section in proton-lead collisions at \texorpdfstring{$\sqrt{s_{_\mathrm{NN}}}=5.02$\TeV}{sqrt(s[NN])=5.02 TeV}}

\author*[kit]{Colin Baus}

\date{\today}

\abstract{
The inelastic hadronic cross section in proton-lead collisions at a centre-of-mass energy per
nucleon pair of 5.02\TeV is measured with the CMS detector at the LHC. The data sample,
corresponding to an integrated luminosity of $\lumi=12.6\pm0.4\nbinv$, has been collected with an unbiased
trigger for inclusive particle production. The cross section is obtained from the measured
number of proton-lead collisions with hadronic activity produced in the pseudorapidity ranges
$3<\eta<5$ and/or $-5<\eta<-3$, corrected for photon-induced contributions, experimental acceptance,
and other instrumental effects. The inelastic cross section is measured to be
$\sigmainel(\text{pPb})=2061\pm3\stat\pm34\syst\pm72\lum\unit{mb}$.
Various Monte Carlo generators, commonly used in heavy
ion and cosmic ray physics, are found to reproduce the data within uncertainties. The value of
$\sigmainel(\text{pPb})$ is compatible with that expected from the proton-proton cross section at 5.02\TeV
scaled up within a simple Glauber approach to account for multiple scatterings in the lead nucleus,
indicating that further net nuclear corrections are small.
}

\hypersetup{%
pdfauthor={CMS Collaboration},%
pdftitle={Measurement of the inelastic cross section in proton-lead collisions at a centre-of-mass energy per nucleon pair of 5.02 TeV},%
pdfsubject={CMS},%
pdfkeywords={CMS, forward physics, proton-lead, cross section}}

\maketitle

\section{Introduction}
The measurement of the inelastic cross section in proton-lead
collisions, $\sigmainel(\text{pPb})$, at a centre-of-mass energy per nucleon
pair of 5.02\TeV performed by the CMS experiment at the CERN LHC is
presented.
The inelastic cross section (also called ``particle-production"~\cite{Engel:1998pw}
or ``absorption"~\cite{Carroll:1978hc} cross section in
previous studies) is defined to include all hadronic events, including contributions from diffractive processes,
except those from the quasi-elastic excitation of the
lead nucleus---estimated to amount to about 100\unit{mb} for the pPb system~\cite{Alvioli:2009iw}.
Inelastic electromagnetic (photon-proton) collisions are also excluded from the measurement.

While being one of the most inclusive
observables in hadronic collisions, the inelastic cross
section is one of the least theoretically accessible quantities, as it cannot be determined from
first-principles calculations of the theory of the strong interaction, quantum chromodynamics.
In proton-proton (pp) and nucleus-nucleus collisions at the LHC,
particles produced in hadronic interactions come mostly from the
hadronisation of quarks and gluons, either produced in semi-hard
scatterings (``minijets'')~\cite{Sarcevic:1988tu} or emitted at very forward
rapidities from "spectator" partons, as well as from
soft diffractive processes in "peripheral" interactions.
From the measured inelastic proton-proton (or nucleon-nucleon) cross section at a given collision
energy, one can theoretically derive the corresponding proton-nucleus and nucleus-nucleus cross
sections by means of Glauber~\cite{Glauber:1955qq,Glauber:1970jm} or Gribov--Regge~\cite{Gribov67a-e}
multiple-scattering approaches that take
into account the known transverse matter profile of nuclei. Key quantities for the experimental
comparison between nucleus-nucleus and pp collisions---such as the nuclear overlap function, the number of
nucleon-nucleon collisions and of par\-tici\-pant nucleons~\cite{d'Enterria:2003qs,Miller:2007ri}---are
also commonly computed through such
approaches. Validating the Glauber and Gribov-Regge predictions with proton-nucleus collisions at
LHC energies has important implications beyond collider physics. Such approaches constitute
crucial ingredients in the Monte Carlo modelling of cosmic ray
air showers at the highest energies~\cite{dEnterria:2011kw}, for which the inelastic cross sections measured in the laboratory
must be extrapolated over a wide energy range. In fact, the inelastic proton-air (mostly proton-nitrogen and
proton-oxygen) cross section
introduces one of the largest uncertainties
for air shower simulations~\cite{Ulrich:2010rg,Parsons:2011ad}.

The Glauber multiple-collision model, based on the eikonal limit (\ie straight-line trajectories of
the colliding nucleons), is the simplest and most economical approach often used to derive inclusive
proton-nucleus quantities from the pp cross sections and, vice versa, to obtain pp cross sections from the cosmic ray
measurements~\cite{Collaboration:2012wt}. However, some of the
approximations applied in the model---foremost the absence of short-range nucleon correlations~\cite{Frankfurt:1993sp} and of inelastic
screening~\cite{Good:1960ba}---impact the computed cross section values. This is observed for fixed-target
proton-carbon data~\cite{Dersch:1999zg,Bellettini:1966zz,Murthy:1975qb,Engler:1970dt,Babaev:1973ff}
and estimated for collider~\cite{CiofidegliAtti:2011fh,Alvioli:2012qa} as well as
ultra-high cosmic ray~\cite{Collaboration:2012wt} energies,
where corrections to the proton-air cross section of the order of 10\% have been obtained.
Short-range correlations increase the number of nucleon-nucleon collisions at small
impact parameters yielding a larger nucleus-nucleus cross section. On the other hand, screening affects the
number of nucleons that are diffractively excited in the multiple collisions but revert back to
their ground state before the scattering process is completed, thereby reducing the nuclear cross
section. Different implementations of such effects exist in the current hadronic interaction
models~\cite{Good:1960ba,Kalmykov:1993qe,Guzey:2005tk,Werner:2005jf,Ostapchenko:2010gt,Ostapchenko:2010vb,Frankfurt:1993qi,Harrington:2002gc}.
A measurement of \sigmainel in pPb collisions
at the LHC can test if the precision of the standard Glauber
calculation is sufficient, and at which energies
corrections to the Glauber approach may become relevant.

\section{Experimental setup and Monte Carlo simulations}

The measurement presented here is based on pPb data taken with the
CMS experiment at the LHC at the beginning of 2013.
A detailed description of the apparatus can be found in~\cite{Chatrchyan:2008zzk}. The main
detector used in this analysis is the hadron forward (HF) calorimeter
that covers the pseudorapidity interval $3<\abs{\eta}<5$. The
calorimeter is composed of quartz fibres in a steel matrix with a
$0.175{\times}0.175$ segmentation in the azimuthal angle $\phi$ (in radians) and
pseudorapidity $\eta$. The quartz fibres pick up the Cherenkov light
produced by the charged component of showers. This light is then
measured by photodetector tubes. The hadronic and electromagnetic
signals of each segment, as derived from fibres of two different lengths and depths,
are combined to form a \emph{tower} signal.

The data used in this analysis comprise an integrated luminosity of
$\lumi=12.6\pm0.4\nbinv$. This dataset combines the integrated luminosities of the two possible directions of the proton and lead beams:
$5.0\pm0.2\nbinv$ and $7.6\pm0.3\nbinv$, for the proton beam going respectively in the clockwise
(negative $\eta$) and anticlockwise (positive $\eta$) direction.
The events are collected using an unbiased trigger, only
requiring the presence of both beams in the interaction point, as determined by the
``Beam Pickup Timing for the eXperiments'' (BPTX) devices.
Detector noise is studied
with events that are randomly read out in the absence
of both beams in the detector.
The luminosity determination technique was calibrated
by means of a van der Meer scan~\cite{vanderMeer:296752} for
both beam directions independently, with an uncertainty of 3.5\%~\cite{CMS-PAS-LUM-13-002}.

A Monte Carlo event simulation based on a \GEANTfour detector description~\cite{Agostinelli:2002hh}
is used to model the experimental response and
derive the reconstruction efficiencies. Different event generators are used to simulate hadronic
proton-nucleus collisions. Three models are based on the Gribov-Regge formalism: \DPMJETWITHVER~\cite{Bopp:2005cr},
\EPOSWITHVER~\cite{Werner:2005jf}, and \QGSJETWITHVER~\cite{Ostapchenko:2010gt}; and a fourth one is based on a minijet+Glauber approach:
\HIJINGWITHVER~\cite{Wang:1991hta}. In addition, particle production from photon-proton ($\gamma\Pp$) interactions in
``ultraperipheral" collisions, at impact parameters larger than the sum of proton and lead radii,
needs to be taken into account~\cite{Baltz:2007kq}. Given the large Pb ion charge, and the associated large "equivalent
photon flux" of its electromagnetic field~\cite{Baltz:2007kq}, inelastic photon-proton
collisions result in a non-negligible particle production contribution.
Pure
photon-photon interactions, mostly producing exclusive electron-positron pairs, and photon-nucleus interactions
(where the photon emitted from the proton collides with the Pb ion) have orders-of-magnitude smaller visible
cross sections and are neglected. Photon-proton processes are generated with the \STARLIGHT programme~\cite{Djuvsland:2010qs}
combined either with \DPMJETWITHVERSL or \PYTHIAWITHVERSL~\cite{Sjostrand:2006za}.

\section{Event selection and analysis}
In this analysis three types of cross sections are measured: (i)~$\sigmaobs$ after removal of noise and correction for pileup, (ii)~$\sigmahadlvl$
after further removal of electromagnetic contributions and translation
into a hadron-level quantity, and (iii)~$\sigmainel$ including the
final extrapolation to the total inelastic hadronic cross section.
Two different approaches are used to determine the
number of inelastic events: (1)~a \textbf{single-arm} event
selection that requires a localised calorimetric energy signal above a given threshold in the
HF detector either at positive or negative pseudorapidities, and (2)~a \textbf{double-arm} event selection that requires a localised signal above threshold in both
HF detectors.
The advantage of using these two event selections is that they have very
different sensitivities to diffractive and photon-proton events as well as to detector noise.
Denoting by $E_{\text{HF+}}$ ($E_{\text{HF-}}$) the highest energy measured in an HF tower at positive (negative)
pseudorapidity, an event is tagged as a candidate for an inelastic collision if it has
 a value of
\begin{equation}
\ehf=
  \begin{cases}
   \text{max}(E_{\text{HF+}},E_{\text{HF-}}) &\text{for single-arm selection}\\
   \text{min}(E_{\text{HF+}},E_{\text{HF-}}) &\text{for double-arm selection}\\
  \end{cases}
\label{eq:selcriteria}
\end{equation}
above a given threshold.

\begin{figure}[hbtp]
\centering
\includegraphics[width=.48\textwidth]{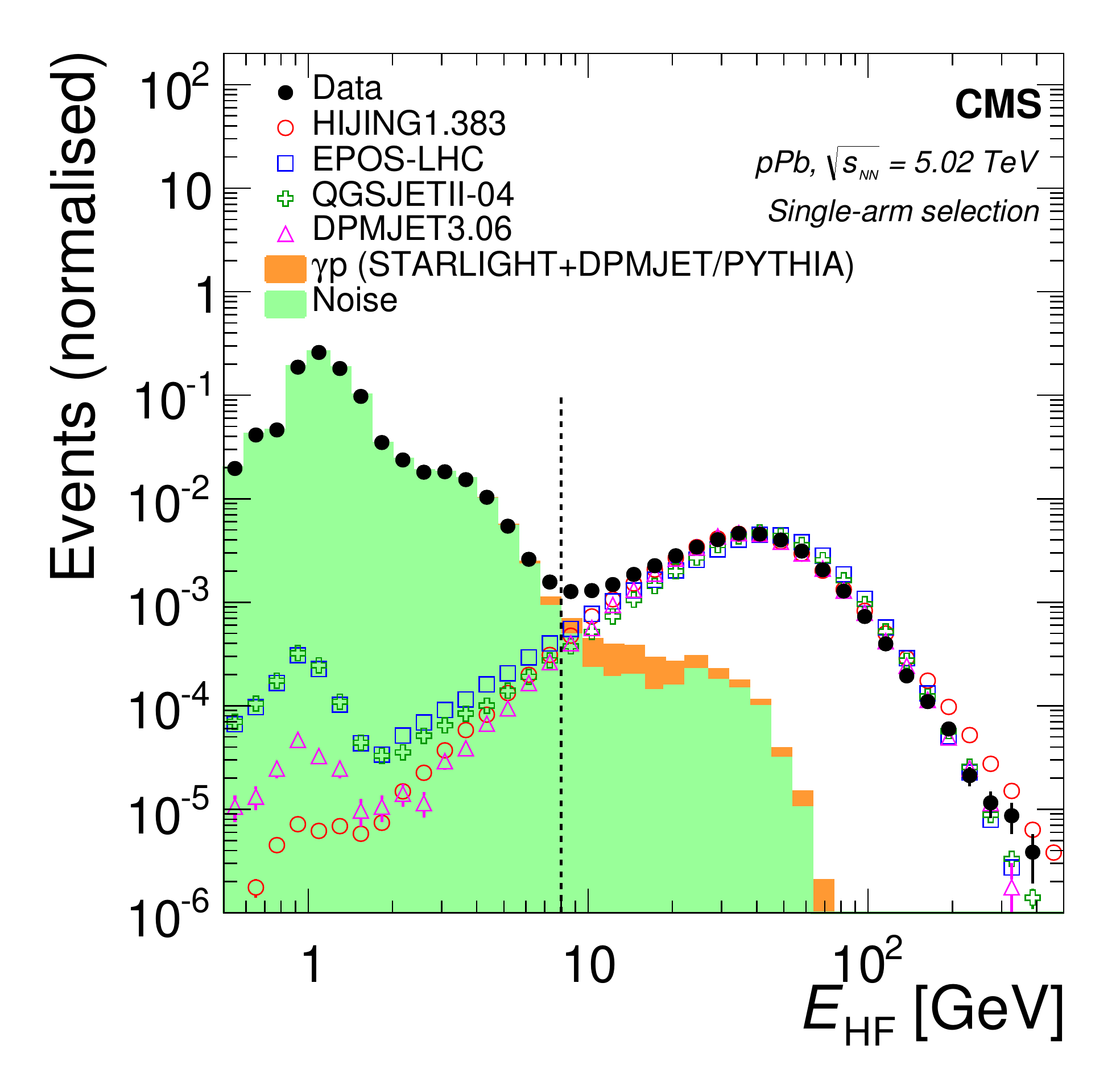}
\includegraphics[width=.48\textwidth]{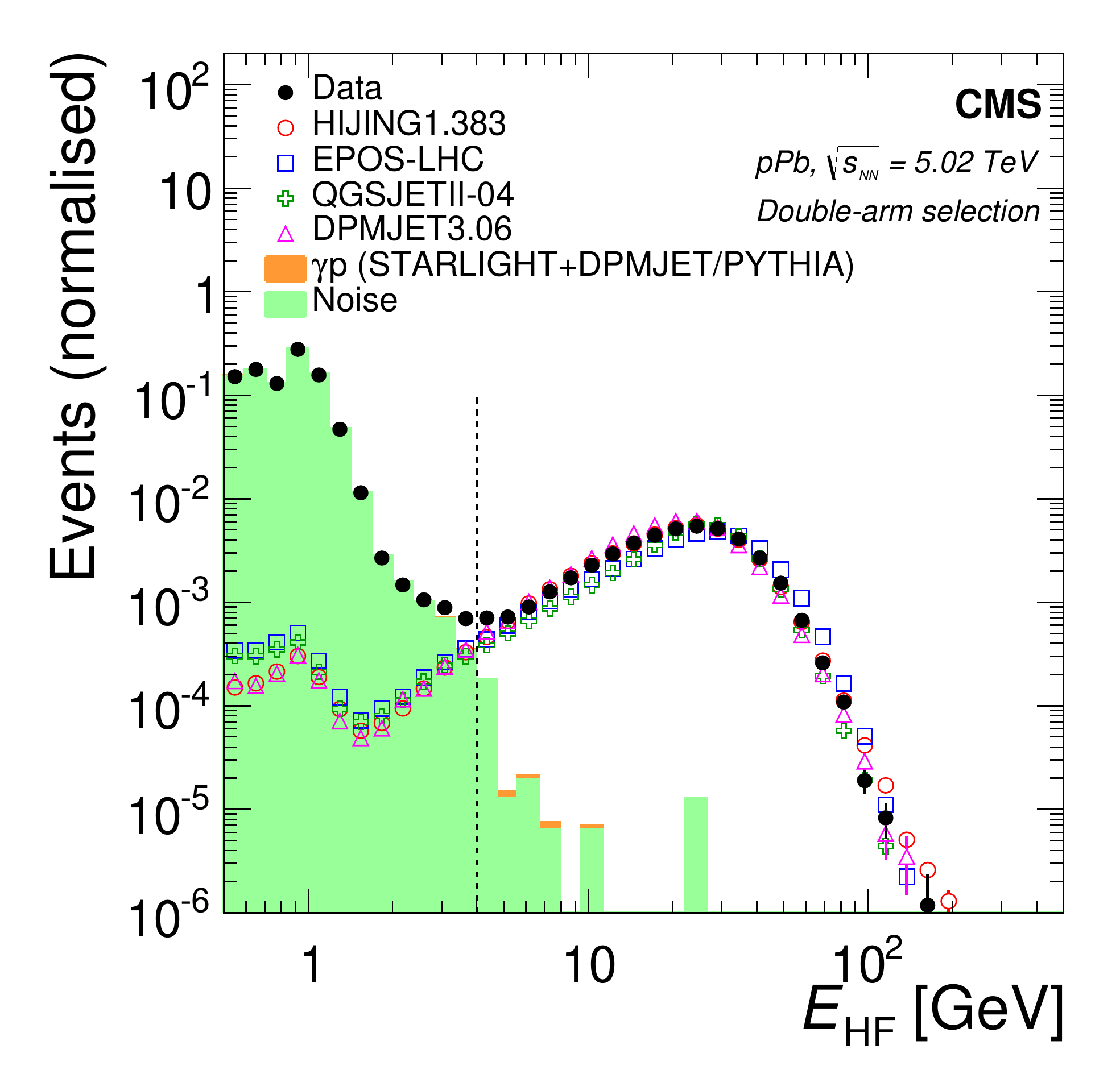}
\caption{
Distribution of the energy deposited in the HF calorimeter (\ehf) for the single-arm (\cmsLeft) and double-arm (\cmsRight) event
selections. The data sample, shown exemplarily for one period with stable run conditions, comprises 1.31\nbinv recorded with an unbiased trigger.
The contribution from noise is
obtained from a random trigger normalised to the same number of triggers as
that in the collision data. The average number of photon-proton processes simulated with
{\STARLIGHT}+\DPMJET and {\STARLIGHT}+\PYTHIA is treated as
background and stacked on top. Four hadronic interaction models (\EPOS, \DPMJET, \HIJING,
and \QGSJET) are overlaid and normalised to the number
of data events with $\ehf>10\GeV$, where the contribution from the
background is small. The vertical line represents the threshold energy
of 8\GeV (4\GeV) for the single-arm (double-arm) selection used in this analysis.}
\label{fig:hfsignal}
\end{figure}

The observed distribution of \ehf is well reproduced by the combined
hadronic inelastic, photon-proton, and detector noise contributions as shown in
the \cmsLeft (\cmsRight) panel of Fig.~\ref{fig:hfsignal} for the single-arm (double-arm)
selection.
The size of the various contributions to the
HF energy deposition is determined from data and simulations. The signal is identified as that coming from
hadronic collisions whereas the backgrounds arise from electromagnetic photon-proton interactions and detector
noise.
The expected number of photon-proton collisions is $\nem=f_{\gamma\Pp}\sigma_{\gamma\Pp}\lumi$,
where $f_{\gamma\Pp}$ is
the fraction of simulated photon-proton
events passing the selection and $\sigma_{\gamma\Pp}$ is the
predicted \STARLIGHT cross section.
The number of misidentified events produced by electronic noise in the
detector is $\nnoise=N f_\text{noise}$, where $f_\text{noise}$ is the fraction of
events read out randomly in the absence of beams that pass the
selection criteria, and $N$ is the number of events recorded with the unbiased trigger.
The estimate of \nnoise includes $N_{\text{obs+noise}}=\nobs f_\text{noise}$
events that contain also an observed inelastic collision, where \nobs
is the number of observed inelastic events.
The double-counted events are explicitly subtracted from \nnoise.
The uncertainty on \nnoise is derived
from variations in different data-taking periods.
The background induced by beam-gas collisions is found to be
negligible deduced from the fraction of events selected with the
trigger indicating the presence of a single beam in the interaction point.

Of the number of inelastic hadronic collisions, $\nhadnocut$,
the ones that are observed by the detector and pass the event selection
are defined as $\nhad$. The purity of the event selection is
$\nhad/\left(\nhad+\nem+\nnoise\right)$,
and the acceptance is given by the ratio
$\epsilon_{\text{acc}}=\nhad/\nhadnocut$.
Both the purity and the acceptance depend on the energy threshold used
for the selection. Higher purity is achieved for the double-arm
selection, since photon-proton interactions lead to a typical final
state where most of the secondary products are asymmetrically emitted
towards the direction of the proton beam. Noise events are also
suppressed by the coincidence requirement. The acceptance is in
general smaller for the double-arm selection due to the smaller chance
of selecting diffractive events characterised by large rapidity gaps
devoid of activity in one or both HF sides.

\begin{figure}[hbtp]
\centering
\includegraphics[width=\cmsFigWidth]{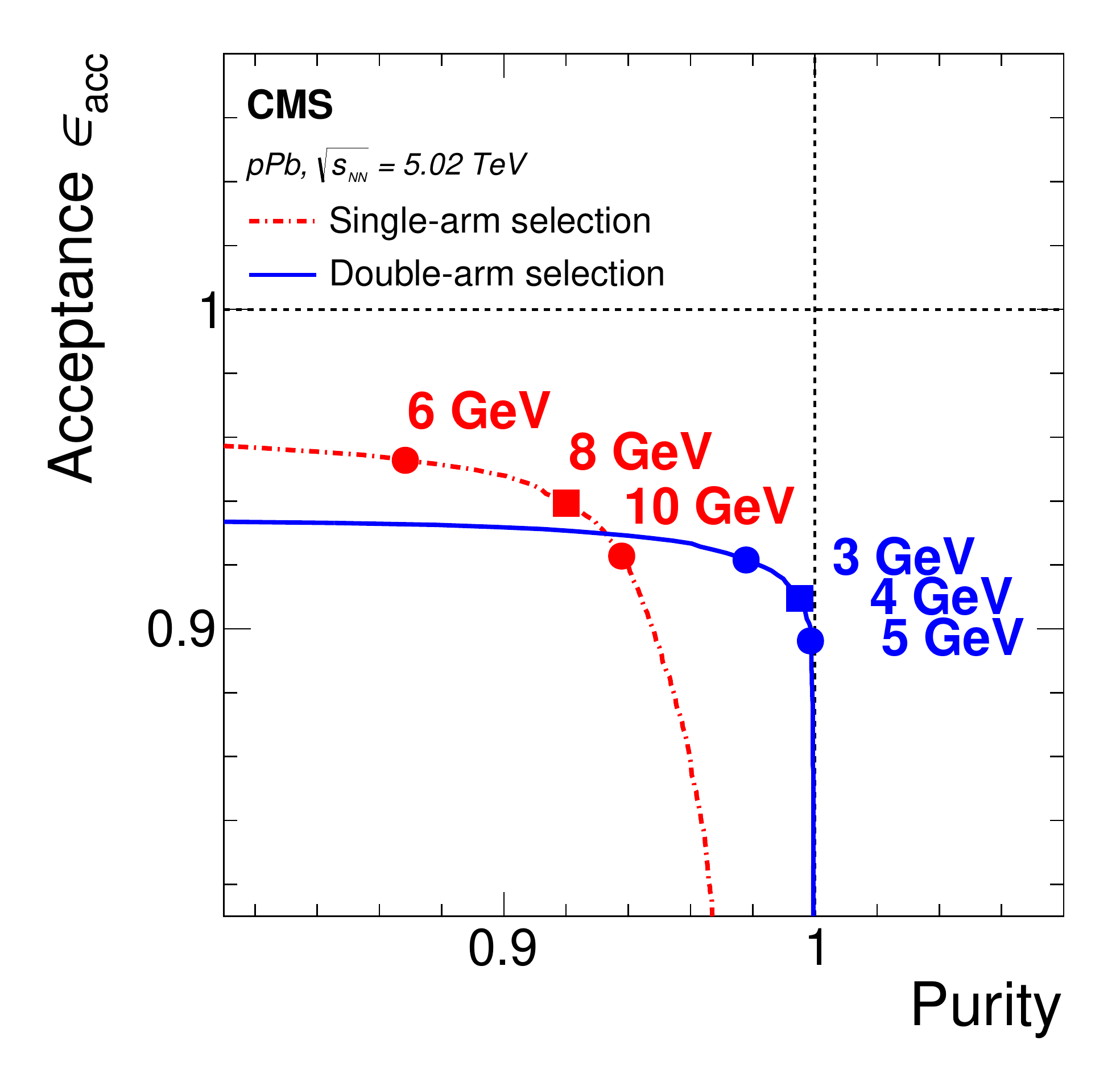}
\caption{Acceptance
versus purity of the two event selections, as derived
from the \EPOS and \QGSJET generators. The symbols indicate
different values of the \ehf thresholds. The chosen thresholds are marked with squares.}
\label{fig:purvseff}
\end{figure}

The dependence of $\epsilon_{\text{acc}}$ on the HF tower energy
threshold is shown in Fig.~\ref{fig:purvseff}. For the single-arm
selection the working point is chosen to be $\ehf>8$\GeV. This value
is the result of a compromise between acceptance (about 93--94\%) and
contamination, while the probability to have a tower above the
threshold does not depend much on the beam direction. The double-arm
selection uses $\ehf>4$\GeV yielding 99\% purity and 91\% acceptance.
The value $\epsilon_{\text{acc}}$ for a specific \ehf threshold
is determined by averaging over the results of the \EPOS
and \QGSJET models. The results of \HIJING and \DPMJET, which do not
include nuclear effects for diffraction, are not considered for this purpose.
Indeed, we have verified that both latter models are unable to describe
the very-forward energy spectra measured with the CASTOR detector ($-6.6<\eta<-5.2$) in events with large rapidity gaps, which are particularly sensitive
to diffractive interactions.

The uncertainties on the $\epsilon_{\text{acc}}$ and \nem values are estimated from the maximum absolute differences
obtained from the results of different event generators, averaged over a wide \ehf interval between 2 and 10\GeV.
The uncertainties on $\epsilon_{\text{acc}}$ are 0.005\,(0.014) and of $\nem/\lumi$ are 11\unit{mb} (0.05\unit{mb}) for the single-arm (double-arm) event selections.

In this analysis no vertex reconstruction is performed and
the impact of contributions from additional pileup (PU) collisions recorded in any given event
is consistently evaluated with the HF detector.
The number of simultaneous collisions is Poisson-distributed
with an expectation value corresponding to the interaction
probability $\lambda$.
If one collision is selected with probability
$\epsilon_{\text{acc}}$, then $i$ simultaneous collisions are selected
with probability $P_i\approx1-(1-\epsilon_{\text{acc}})^i$.
The approximation assumed in the equation, which does not account for
energy deposits of multiple events in the HF towers, was verified to
be valid by means of a toy Monte Carlo simulation. The number of collisions is
then corrected with the factor
$f_{\text{PU}}=\epsilon_{\text{acc}}\lambda/\sum\limits_{i=1}^{\infty}
P_i \; \text{Poisson}(i;\lambda)$.
The interaction probability $\lambda$, which amounts to 2--8\% depending
on the data-taking period, is calculated recursively from the ratio of
the number of inelastic events to the number of unbiased triggers.
The pileup correction increases
the measured cross section by 2\% for both event selections,
and introduces an uncertainty on the final pPb cross section that is smaller than 0.1\%.

To facilitate the direct comparison of the results to model
predictions, detector level quantities, such as \ehf, are translated to
hadron-level quantities. For this purpose, $p_{\mathrm{HF}}$ is defined
equivalently to Eq.~\eqref{eq:selcriteria} but replacing \ehf by the
largest absolute value among the momenta, $\abs{\vec{p}}$, of all
generated final-state particles (with lifetimes above 1\cm/$c$),
within the pseudorapidity intervals of the HF calorimeters
($3<\abs{\eta}<5$), excluding muons and neutrinos.
A correction factor \chad, obtained from simulations, is used to translate the measured cross section into a hadron-level quantity, defined by
the ratio of the number of visible events, which fulfil a given requirement on $p_{\mathrm{HF}}$, to the number of observed events, which pass the selection on \ehf.
Thus, \chad is larger than unity for requiring $p_{\mathrm{HF}}>0$, but will approach zero for very high thresholds.
The threshold can
be chosen freely, and for the present analysis the requirement on the
minimal value of $p_{\mathrm{HF}}$ is chosen such that the fractions of events
passing this selection and passing that on \ehf are equal.
The factor \chad then becomes equal to unity and has no
numerical effect on the central value of the derived cross section.
This procedure leads to the choice of selecting events that
fulfil the requirement $p_{\mathrm{HF}}>21.3\GeVc$ (11.3\GeVc)
for the single-arm (double-arm) analysis.
For the chosen thresholds, the mean of the
\chad values of all four hadronic interaction models is
unity and the slight dependence on models is taken into account as
a systematic uncertainty on \chad equal to the
standard deviation of the four values.

The values of the acceptance, backgrounds, and correction factors
are summarised in Table~\ref{tab:collectedcorr}.

\begin{table*}[!ht]
  \centering
  \topcaption{Central values and uncertainties for the two event selections for
    noise cross section contribution ($\nnoise/\lumi$) and the fraction
    of noise events ($f_{\text{noise}}$) as derived from data. Additionally, the
    quantities acceptance ($\epsilon_{\text{acc}}$), electromagnetic cross section contribution
    ($\nem/\lumi$), and hadron-level correction factor (\chad) as derived from simulations are listed.}
\begin{tabular}{lccccc}
\hline
Selection  & $\nnoise/\lumi$\,[mb] & $f_{\text{noise}}$           & $\epsilon_{\text{acc}}$ & $\nem/\lumi$\,[mb] & \chad\\
\hline
Single-arm & $102\pm25$                   & $(2.0\pm0.5)\times10^{-3}$ & $0.939\pm 0.005$     & $63\pm11$                 & $1.000\pm0.004$\\
Double-arm & $9\pm3$                      & $(1.8\pm0.8)\times10^{-4}$ & $0.910\pm 0.014$     & $0.33\pm0.05$             & $1.000\pm0.002$\\
\hline
\end{tabular}
\label{tab:collectedcorr}
\end{table*}

The number of observed inelastic events, \nobs, is derived from the number of events passing
the event selection, $N_{\text{sel}}$, and is corrected for noise (\nnoise),
double counting ($N_{\text{obs+noise}}$), and pileup ($f_{\text{PU}}$).
Dividing this number by the integrated luminosity yields the \emph{observed} cross section:
\begin{equation}
  \sigmaobs=\frac{\nobs}{\lumi}=\left(N_{\text{sel}}-\nnoise+N_\text{obs+noise}\right) \; \frac{f_{\text{PU}}}{\lumi}.
\end{equation}

Using the relation $N_{\text{obs+noise}}=\nobs f_\text{noise}$ one obtains
\begin{equation}\label{eq:viscorrections}
  \sigmaobs=\frac{1}{\lumi}\frac{N_{\text{sel}}-\nnoise}{1/f_{\text{PU}}-f_{\text{noise}}}.
\end{equation}

The \emph{visible} cross section for hadronic collisions is derived by subtracting
the photon-proton contamination and applying the correction factor
\chad. Its numerical value is, by definition, equal to the
part of the observed cross section related to hadronic collisions:
\begin{equation}\label{eq:hadcorrections}
  \sigmahadlvl=\frac{1}{\lumi}\frac{N_{\text{sel}}-\nnoise-\nem}{1/f_{\text{PU}}-f_{\text{noise}}}\chad.
\end{equation}

The \emph{inelastic} cross section is obtained by
correcting for the limited detector
acceptance ($\epsilon_{\text{acc}}$):
\begin{equation}\label{eq:corrections}
  \sigmainel=\frac{1}{\lumi}\frac{N_{\text{sel}}-\nnoise-\nem}{1/f_{\text{PU}}-f_{\text{noise}}}\frac{1}{\epsilon_{\text{acc}}}.
\end{equation}

The ratio of the visible hadronic cross section obtained with the single-arm selection to the one obtained with the
double-arm selection is sensitive to the fraction of diffractive pPb
events.
The measured value of this ratio allows one to constrain the
diffractive cross section, $\sigma_{\text{diff}}$, in the models. In
order to be compatible within 2 standard deviations of the data, the
\EPOS diffractive cross section cannot be scaled up or down by more
than $\pm$13\% from its default value, while for \QGSJET those limits
are $\pm$20\%.
This propagates into an $\epsilon_{\text{acc}}(\sigma_{\text{diff}})$
uncertainty on \sigmainel, conservatively assumed to be symmetric, of
0.8\%\,(1.1\%). For this and the following uncertainties,
the first number is related to the single-arm selection and the bracketed one to the double-arm selection.
The model-dependence of the
acceptance corrections results in an uncertainty for
$\epsilon_{\text{acc}}(\text{models})$ of 0.5\%\,(1.6\%) for the two
selections, respectively.

Since less than half of the diffractive events, mostly with a
high-mass diffractive system, pass the hadron-level selection, the
uncertainty on \chad is smaller than that on
$\epsilon_{\text{acc}}$.
The 1 standard deviation differences found among the four hadronic interaction models
on the hadron-level correction, \chad,
propagate into uncertainties
on \sigmahadlvl of 0.4\%\,(0.2\%) for the single-arm (double-arm) selection.
The subtraction of photon-proton events (with the \nem uncertainty shown in Table~\ref{tab:collectedcorr}),
results in an uncertainty of $0.6\%\,({<}0.1\%)$
on \sigmainel and \sigmahadlvl. The uncertainty on \nnoise propagates into a
1.3\%\,(0.2\%) uncertainty in the final cross sections.
The effect on the event selection of the radiation damage in the HF fibres is assessed by
rescaling the signals of the simulated HF response to match data
in segments of pseudorapidity.
The rescaling factors are calculated using the average response produced by
\EPOS, \HIJING, and \QGSJET.
These scaling factors are found to be consistent with the observed radiation damage of HF
and range from 1 to 0.67, depending on pseudorapidity.
The amount of radiation damage is estimated from a comparison of $\rd E/\rd\eta$ distributions measured in proton-proton
collisions at $\sqrt{s}=2.76\,$\TeV recorded in 2013 and in 2010.
The systematic
uncertainty induced on the cross section by this approach is estimated by repeating the measurement
without the radiation damage correction, which introduces an
effect of 1.7\%\,(0.8\%) on the cross section.
As a further check of the HF tower
energy resolution, the cross sections are
computed by increasing the selection thresholds to $\ehf>10\GeV\,(5\GeV)$. To account for
both effects, a systematic uncertainty on the cross section of 0.6\%\,(0.4\%) is added.
The cross sections measured for the two beam directions are found to be consistent.
Consequently, no dedicated systematic uncertainty is assigned to this effect.

\begin{table*}[!htb]
  \centering
  \topcaption{List of the systematic uncertainties, propagated into the final pPb cross sections, for the two event selections.}
\begin{tabular}{llcc}
\hline
  Source of uncertainty && Single-arm & Double-arm \\ \hline

  Noise subtraction (\nnoise) && 1.3\% & 0.2\% \\
  Pileup correction ($f_{\text{PU}}$) && ${<}$0.1\% & ${<}$0.1\% \\
  Acceptance ($\epsilon_{\text{acc}}(\text{models})$) && 0.5\% & 1.6\% \\
  Acceptance ($\epsilon_{\text{acc}}(\sigma_{\text{diff}})$) && 0.8\% & 1.1\% \\
  Hadron-level correction (\chad) && 0.4\% & 0.2\% \\
  Photon-proton subtraction (\nem) && 0.6\% & ${<}$0.1\% \\
  Detector simulation && 1.7\% & 0.8\% \\
  HF energy thresholds && 0.6\% & 0.4\% \\
  Integrated luminosity (\lumi) && 3.5\% & 3.5\% \\ \hline
\end{tabular}
\label{tab:systematics}
\end{table*}

All the different sources of uncertainty of the measurement are listed
in Table~\ref{tab:systematics} for the single-arm and double-arm event
selections. The three derived cross sections have different systematic
uncertainties since not all contributions are relevant to each of
them. For \sigmainel, all uncertainties but the one due to the
hadron-level correction contribute. The total uncorrelated systematic
uncertainty is therefore 2.5\%\,(2.2\%) for the single-arm
(double-arm) selection.  For \sigmahadlvl, the dominant uncertainty is
due to the hadron-level correction instead of the correction for
$\epsilon_{\text{acc}}$. The value of the uncertainty is therefore
reduced to 2.3\%\,(0.9\%). The uncertainties for detector simulation
and photon-proton correction do not contribute to \sigmaobs and,
hence, its uncertainty becomes 1.4\%\,(0.5\%). For all cross sections,
a (dominant) integrated luminosity uncertainty of 3.5\% is added. The
main contributions to the latter arise from the model used to
describe the beam profile, the length scale of the beam displacement,
and the bunch-to-bunch variations~\cite{CMS-PAS-LUM-13-002}.

\section{Results and summary}

The measured cross sections for the single-arm and double-arm event selections are listed in
Table~\ref{tab:viscs}, compared to the predictions of the hadronic interaction models \DPMJET, \EPOS, and \QGSJET.
Due to the different acceptance, the
extrapolations from the hadron-level to the inelastic cross section are of different
magnitude, but the models reproduce well the approximately 65\unit{mb} difference between
the two selections. The values of the
inelastic cross sections obtained from the single-arm and double-arm methods
agree well within the uncertainties.

\begin{table*}[!htb]
  \centering
  \topcaption{Summary of cross sections obtained from the two different event selections. The acceptance definition for \sigmahadlvl is based on the production of stable particles within $3<\abs{\eta}<5$ with momentum $p_{\rm HF}>21.3\GeVc$ (11.3\GeVc) for the single-arm (double-arm) event selections.}
\begin{tabular}{llccc}
\hline
  & Selection & \sigmaobs\,(mb) & \sigmahadlvl\,(mb) & \sigmainel\,(mb)\\ \hline
  \multirow{2}{*}{Data}           & Single-arm & 2003$\pm$76 & 1937$\pm$82 & 2063$\pm$89\\
                                  & Double-arm & 1873$\pm$66 & 1872$\pm$68 & 2059$\pm$85\\ \hline
  \multirow{2}{*}{\EPOSWITHVER}   & Single-arm & \NA & 1947 & \multirow{2}{*}{2082}\\
                                  & Double-arm & \NA & 1883 & \\ \hline
  \multirow{2}{*}{\QGSJETWITHVER} & Single-arm & \NA & 2059 & \multirow{2}{*}{2181}\\
                                  & Double-arm & \NA & 1998 & \\ \hline
  \multirow{2}{*}{\DPMJETWITHVER} & Single-arm & \NA & 2116 & \multirow{2}{*}{2166}\\
                                  & Double-arm & \NA & 2055 & \\ \hline
\end{tabular}
\label{tab:viscs}
\end{table*}

The final \sigmainel value is obtained by taking the weighted average of the
measured values in the two event selections.
The statistical uncertainties and the uncertainty
on the luminosity are correlated between the selections.
The degree of correlation among the remaining systematic uncertainties is much
smaller and they are taken as uncorrelated.
This yields a final result for the inelastic hadronic cross section of
\begin{equation*}
\label{eq:result}
\sigmainel(\text{pPb})=2061\pm3\stat\pm34\syst\pm72\lum\unit{mb}.
\end{equation*}

\begin{figure}[hbtp]
\centering
\includegraphics[width=\cmsFigWidth]{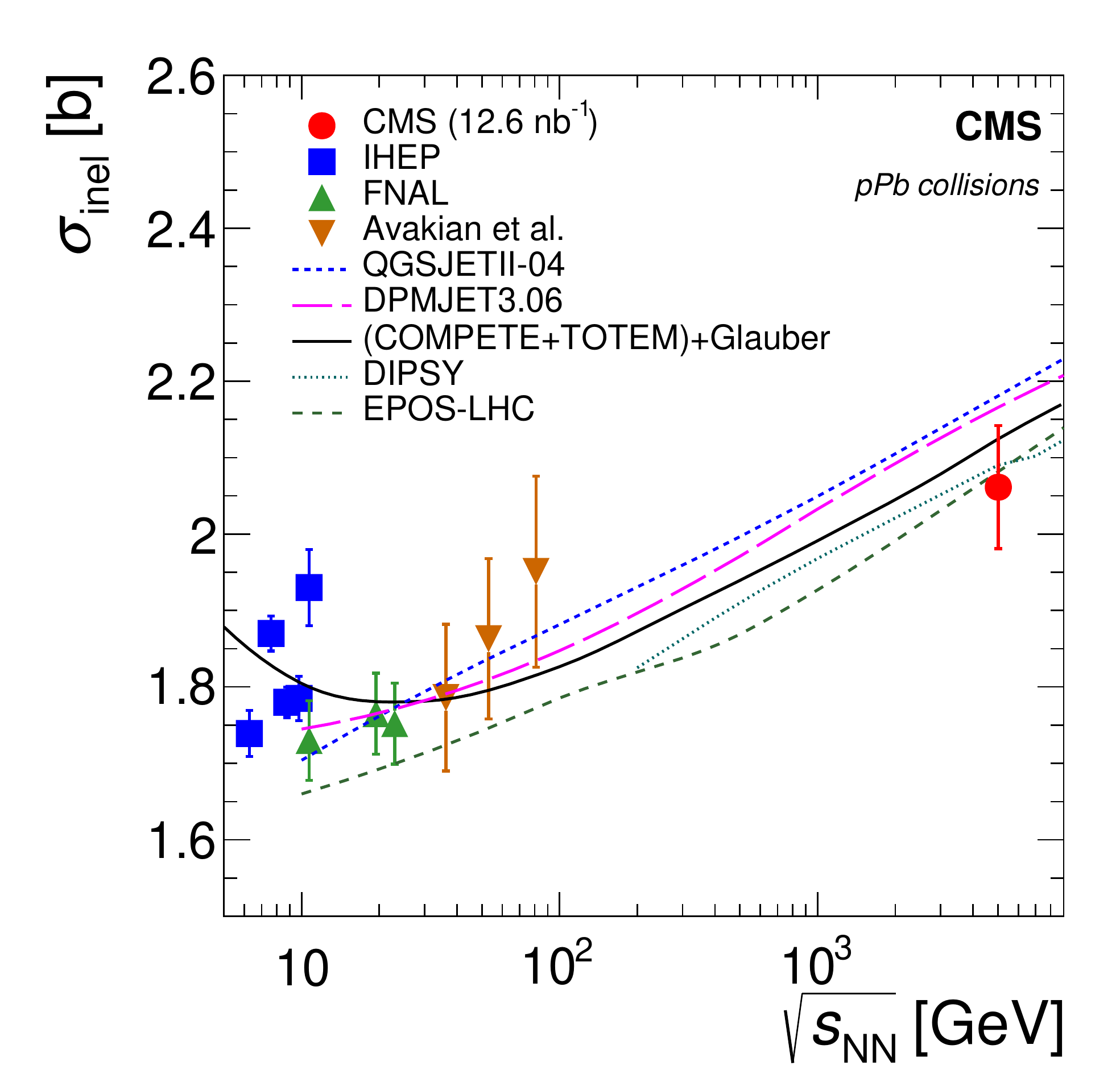}
\caption{Inelastic hadronic cross sections for pPb collisions as a
  function of the centre-of-mass energy. The measurement described
  here (circle, with error bars obtained from the quadratic sum of all
  uncertainties) is compared to lower energy
  data (squares and triangles)~\cite{Denisov:1973zv,Carroll:1978hc,Avakian:1986ek} and to
  different model predictions (curves).}
\label{fig:conclusion2}
\end{figure}

This result is shown in Fig.~\ref{fig:conclusion2} compared to other
measurements at different centre-of-mass energies and to various
theoretical predictions. A pPb cross section was also measured by the
ALICE Collaboration, amounting to 2090--2120\unit{mb} with an
uncertainty of 70\unit{mb}~\cite{Abelev:2014epa}.
A direct comparison of this observed cross section to the one measured in the
present analysis is not possible due to the unknown to us ALICE detector
acceptance and possible contamination from noise and photon-proton interactions.

The inelastic cross section measured by the CMS experiment is
compared to the Glauber-model prediction (solid curve in Fig.~\ref{fig:conclusion2}) obtained using
a pp inelastic cross section at
$\sqrt{s}=5.02\TeV$ of $70.0\pm1.5$\unit{mb}, derived from the COMPETE parametrisation~\cite{Cudell:2002xe} including the
measurement of the TOTEM Collaboration at $\sqrt{s}=7$\TeV~\cite{Antchev:2013iaa} (where the assigned uncertainty is
that measured by the latter). The Glauber calculation yields $2130\pm40$\unit{mb} and is compatible with the measurement
presented here indicating that effects neglected by the calculation (such as nucleon correlations and screening) are
either small or approximately cancel out.
The experimental result is also consistent with the
prediction of the \DIPSY model~\cite{Avsar:2005iz,Flensburg:2011kk}
based on a dipole-model approach including parton saturation and multiple-scattering.
Among the Gribov--Regge models, the \EPOS prediction is compatible with the measurement within
uncertainties, whereas \DPMJET and \QGSJET predict a value more than 1 standard deviation above the
data, with a larger discrepancy appearing for the \sigmahadlvl cross
sections (Table~\ref{tab:viscs}).
The \EPOS and \QGSJET models are commonly used for cosmic ray air shower
simulations. Thus, at the corresponding cosmic ray proton energies of
$E_\mathrm{cr}=s/(2m_{\Pp})=10^{16.1}$\unit{eV}, where $m_{\Pp}$ is the
mass of the proton, there are no indications for data-model deviations
above $\approx$5\% in the proton-lead collisions studied here.
Note, however, that our measurement deals with an ion much
heavier than those involved in proton-air interactions.
Corrections to the Glauber model are possibly larger in
the latter case~\cite{Alvioli:2009iw,Collaboration:2012wt}.
In summary, the measurement of the cross sections in pPb collisions
presented here is the first such fully corrected measurement at
multi-TeV energies and, thus, provides important constraints on
hadronic interaction models commonly used in high-energy heavy ion and
cosmic ray physics.

\begin{acknowledgments}
We congratulate our colleagues in the CERN accelerator departments for the excellent performance of the LHC and thank the technical and administrative staffs at CERN and at other CMS institutes for their contributions to the success of the CMS effort. In addition, we gratefully acknowledge the computing centres and personnel of the Worldwide LHC Computing Grid for delivering so effectively the computing infrastructure essential to our analyses. Finally, we acknowledge the enduring support for the construction and operation of the LHC and the CMS detector provided by the following funding agencies: BMWFW and FWF (Austria); FNRS and FWO (Belgium); CNPq, CAPES, FAPERJ, and FAPESP (Brazil); MES (Bulgaria); CERN; CAS, MoST, and NSFC (China); COLCIENCIAS (Colombia); MSES and CSF (Croatia); RPF (Cyprus); MoER, ERC IUT and ERDF (Estonia); Academy of Finland, MEC, and HIP (Finland); CEA and CNRS/IN2P3 (France); BMBF, DFG, and HGF (Germany); GSRT (Greece); OTKA and NIH (Hungary); DAE and DST (India); IPM (Iran); SFI (Ireland); INFN (Italy); MSIP and NRF (Republic of Korea); LAS (Lithuania); MOE and UM (Malaysia); CINVESTAV, CONACYT, SEP, and UASLP-FAI (Mexico); MBIE (New Zealand); PAEC (Pakistan); MSHE and NSC (Poland); FCT (Portugal); JINR (Dubna); MON, RosAtom, RAS and RFBR (Russia); MESTD (Serbia); SEIDI and CPAN (Spain); Swiss Funding Agencies (Switzerland); MST (Taipei); ThEPCenter, IPST, STAR and NSTDA (Thailand); TUBITAK and TAEK (Turkey); NASU and SFFR (Ukraine); STFC (United Kingdom); DOE and NSF (USA).

Individuals have received support from the Marie-Curie programme and the European Research Council and EPLANET (European Union); the Leventis Foundation; the A. P. Sloan Foundation; the Alexander von Humboldt Foundation; the Belgian Federal Science Policy Office; the Fonds pour la Formation \`a la Recherche dans l'Industrie et dans l'Agriculture (FRIA-Belgium); the Agentschap voor Innovatie door Wetenschap en Technologie (IWT-Belgium); the Ministry of Education, Youth and Sports (MEYS) of the Czech Republic; the Council of Science and Industrial Research, India; the HOMING PLUS programme of the Foundation for Polish Science, cofinanced from European Union, Regional Development Fund; the OPUS programme of the National Science Center (Poland); the Compagnia di San Paolo (Torino); the Consorzio per la Fisica (Trieste); MIUR project 20108T4XTM (Italy); the Thalis and Aristeia programmes cofinanced by EU-ESF and the Greek NSRF; the National Priorities Research Program by Qatar National Research Fund; the Rachadapisek Sompot Fund for Postdoctoral Fellowship, Chulalongkorn University (Thailand); and the Welch Foundation, contract C-1845.
\end{acknowledgments}
\bibliography{auto_generated}
\cleardoublepage \appendix\section{The CMS Collaboration \label{app:collab}}\begin{sloppypar}\hyphenpenalty=5000\widowpenalty=500\clubpenalty=5000\textbf{Yerevan Physics Institute,  Yerevan,  Armenia}\\*[0pt]
V.~Khachatryan, A.M.~Sirunyan, A.~Tumasyan
\vskip\cmsinstskip
\textbf{Institut f\"{u}r Hochenergiephysik der OeAW,  Wien,  Austria}\\*[0pt]
W.~Adam, E.~Asilar, T.~Bergauer, J.~Brandstetter, E.~Brondolin, M.~Dragicevic, J.~Er\"{o}, M.~Flechl, M.~Friedl, R.~Fr\"{u}hwirth\cmsAuthorMark{1}, V.M.~Ghete, C.~Hartl, N.~H\"{o}rmann, J.~Hrubec, M.~Jeitler\cmsAuthorMark{1}, V.~Kn\"{u}nz, A.~K\"{o}nig, M.~Krammer\cmsAuthorMark{1}, I.~Kr\"{a}tschmer, D.~Liko, T.~Matsushita, I.~Mikulec, D.~Rabady\cmsAuthorMark{2}, B.~Rahbaran, H.~Rohringer, J.~Schieck\cmsAuthorMark{1}, R.~Sch\"{o}fbeck, J.~Strauss, W.~Treberer-Treberspurg, W.~Waltenberger, C.-E.~Wulz\cmsAuthorMark{1}
% ...
\vskip\cmsinstskip
% ...
1:~~Also at Vienna University of Technology, Vienna, Austria\\
2:~~Also at CERN, European Organization for Nuclear Research, Geneva, Switzerland\\
% ...

\end{sloppypar}
\end{document}